\documentclass[twocolumn]{jpsj3} %% two-column layout
\usepackage{color}
\title{
Anomalous Behavior of the Magnetization Process 
of the $S=1/2$ Kagome-Lattice Heisenberg Antiferromagnet 
at One-Third Height of the Saturation
}
\catcode`\@=11
\def\simle{\mathrel{\mathpalette\@versim<}}   % < over \sim
\def\simge{\mathrel{\mathpalette\@versim>}}   % > over \sim
\def\@versim#1#2{\lower2.5pt\vbox{\baselineskip0pt \lineskip-.5pt
   \ialign{$\m@th#1\hfil##\hfil$\crcr#2\crcr\sim\crcr}}}
\catcode`\@=12

\author{Hiroki Nakano$^{1}$
\thanks{E-mail: hnakano@sci.u-hyogo.ac.jp} 
and 
T\^oru Sakai$^{1,2}$
\thanks{E-mail: sakai@spring8.or.jp} 
}

\inst{$^{1}$Graduate School of Material Science, 
University of Hyogo,
Kamigori, 
Hyogo 678-1297, Japan \\
$^{2}$
Japan Atomic Energy Agency, SPring-8, 
Sayo, Hyogo 679-5148, Japan 
}

\recdate{\today}

\abst{
The magnetization process of the $S=1/2$ Heisenberg 
antiferromagnet on the kagome lattice is 
studied by the numerical-diagonalization method. 
We successfully obtain a new result 
of the magnetization process of a 42-site cluster 
in the entire range. 
Our analysis clarifies that 
the critical behavior around 
one-third of the height of the saturation 
is different from the typical behavior 
of the well-known magnetization plateau 
in two-dimensional systems. 
We also examine the effect of the $\sqrt{3}\times\sqrt{3}$-type 
distortion added to the kagome lattice. 
We find 
at one-third of the height of the saturation 
in the magnetization process 
that the undistorted kagome point 
is just the boundary between two phases 
that show their own properties that are different from each other. 
Our results suggest a relationship between 
the anomalous critical behavior at the undistorted point 
and 
the fact that the undistorted point is the boundary. 
}

\begin{document}
\maketitle

\section{Introduction} 
%% No sections necessary for express letters, letters and short notes

Various characteristics of a magnetic material 
are included in its magnetization process. 
They provide us with useful information 
to understand the properties of a material well. 
Among such characteristic behaviors, 
the phenomenon of magnetization plateaux has long attracted 
the attention of many experimental and theoretical researchers. 
A magnetization plateau is the behavior of the appearance 
of a region of magnetic field in a magnetization process 
where the magnetization does not increase 
even with an increase in the magnetic field, 
in contrast to the fact that a normal magnetization process 
shows a smooth and significant increase in the magnetization 
with an increase in the magnetic field. 
The magnetization plateau originates from the existence 
of an energy gap between states with different magnetizations; 
the nonzero gap occurs owing to the formation 
of an energetically stable quantum spin state. 
The field dependence of magnetization just outside the plateau 
generally originates from the nature of the density of states
determined from the parabolic dispersion 
of states next to the energy gap,  
where the field dependence of magnetization shows 
a characteristic exponent $\delta$ defined in the form of 
\begin{equation}
|m-m_{\rm c}| 
\sim 
|h-h_{\rm c}|^{1/\delta} .
\label{exponent_def}
\end{equation}
Therefore, this exponent is determined 
by the spatial dimension of the system: 
for example, $\delta=1$ for the two-dimensional system. 

Under these circumstances, 
it is an interesting case if 
it has attracted the attention 
of many physicists studying magnetism, 
in terms of whether the magnetization plateau is formed 
and 
how the magnetization behaves as a function of the magnetic field 
outside the plateau, that is, 
the $S=1/2$ Heisenberg antiferromagnet on the kagome lattice. 
In recent years, the kagome-lattice antiferromagnet 
has attracted increasing interest 
not only as a theoretical toy model 
but also from the viewpoint of discoveries 
of several realistic materials: 
herbertsmithite\cite{Shores_herbertsmithite2005,
Mendels_herbertsmithite2010}, 
volborthite\cite{Yoshida_jpsj_volborthite2009,
Yoshida_prl_volborthite2009}, and 
vesignieite\cite{Okamoto_jpsj_vesignieite2009,
Okamoto_prb_vesignieite2011}. 
Since the kagome-lattice antiferromagnet is 
a typical two-dimensional frustrated system, 
most theoretical studies have been carried out 
mainly by the basis of the numerical-diagonalization 
method\cite{Lecheminant,Waldtmann,Hida_kagome,
cabra2002,Honecker0,Honecker1,cabra2005,Cepas,
Spin_gap_Sindzingre,kgm_ramp,Sakai_HN_PRBR,kgm_gap,
Honecker2011,capponi2013}. 
The brief behavior of the magnetization process 
of the kagome-lattice antiferromagnet  
was clarified in Ref.~\ref{Hida_kagome}, 
which showed the existence 
of the magnetization plateau at one-third of the height 
of the saturation. 
This result was supported by Ref.~\ref{Honecker1}. 
However, these studies did not focus much on 
the behavior just outside of the plateau. 
References~\ref{kgm_ramp} and \ref{Sakai_HN_PRBR}, 
on the other hand, pointed out that 
the behavior outside of the state of this height 
is different from 
that of the well-known magnetization plateau 
explained above, and that 
the width of the finite-size step at this height 
possibly vanishes in the thermodynamic limit. 
In particular, Ref.~\ref{Sakai_HN_PRBR} showed 
anomalous critical exponents for the behavior 
just outside of the one-third magnetization state, 
which are different from the exponent $\delta=1$ 
for the typical magnetization plateau 
of a two-dimensional system. 
The authors of Ref.~\ref{Honecker2011}, however, 
considered that 
the width of the plateau survives from the studies 
based on the model with easy-axis exchange 
anisotropies\cite{cabra2002,cabra2005}, 
although the authors did not mention the behavior 
just outside of the plateau. 

Recently, on the other hand, there have been an increasing number 
of studies where the density matrix renormalization group (DMRG) 
calculations are applied to the kagome-lattice 
antiferromagnet\cite{Jiang2008,Yan_Huse_White_DMRG,
Depenbrock2012,SNishimoto_NShibata_CHotta}. 
By the grand canonical analysis\cite{CHotta_NShibata} 
based on their DMRG calculations applied 
to the kagome-lattice antiferromagnet, 
Ref.~\ref{SNishimoto_NShibata_CHotta} reported the existence 
of the magnetization plateau at the one-third height 
together with those at the one-ninth, five-ninth, 
and seven-ninth heights, 
although the authors did not discuss 
the behavior just outside of the plateaux; 
the author of Ref.~\ref{SNishimoto_NShibata_CHotta} 
also proposed quantum spin states with a nine-site structure 
as the states characterizing the plateaux. 
If such a state is stably realized at the one-third height, 
the above argument of the relationship between 
the parabolic dispersion and the critical exponent $\delta$ 
might give a conclusion 
that the upper and lower-side 
dispersions give the normal critical behavior 
in the field dependence of magnetization just outside the plateau. 

The purpose of this paper is to study 
the true behavior of the magnetization process 
for the $S=1/2$ Heisenberg antiferromagnet 
on the kagome lattice 
with as much effort as possible 
on the basis of numerical-diagonalization data. 
We tackle this issue from the following two routes. 
One is to examine our new result 
of the magnetization process for a 42-site cluster. 
To the best of our knowledge, 
this is the first report on the magnetization process 
of this size in the $S=1/2$ model within the entire 
range from the zero magnetization 
to the saturation\cite{comment_size}. 
Note here that this large-scale calculation 
has been carried out 
using the K computer, Kobe, Japan. 
The other route is to examine 
the change that occurs 
by adding a distortion to the ideal undistorted kagome lattice. 
In this study, we investigate 
the case of the $\sqrt{3}\times\sqrt{3}$ type. 
From these results, we try to clarify the reason 
for the discrepancy between the numerical-diagonalization 
and DMRG studies concerning the behavior 
at the one-third height. 

This paper is organized as follows. 
In the next section, the model that we study here is introduced. 
The method and analysis procedure are also explained. 
The third section is devoted to 
the presentation and discussion of our results. 
Our results including new data for the 42-site cluster 
clearly indicate anomalous critical exponents 
for the field dependence of magnetization just outside the plateau. 
Examining the effect of the distortion in the kagome lattice 
clarifies that the undistorted point is just at the boundary 
between two phases, which are different from each other. 
In the final section, we present our conclusion 
together with some remarks and discussion. 

\section{Model Hamiltonians, Method, and Analysis} 

\begin{figure}[htb]
\begin{center}
\includegraphics[width=7cm]{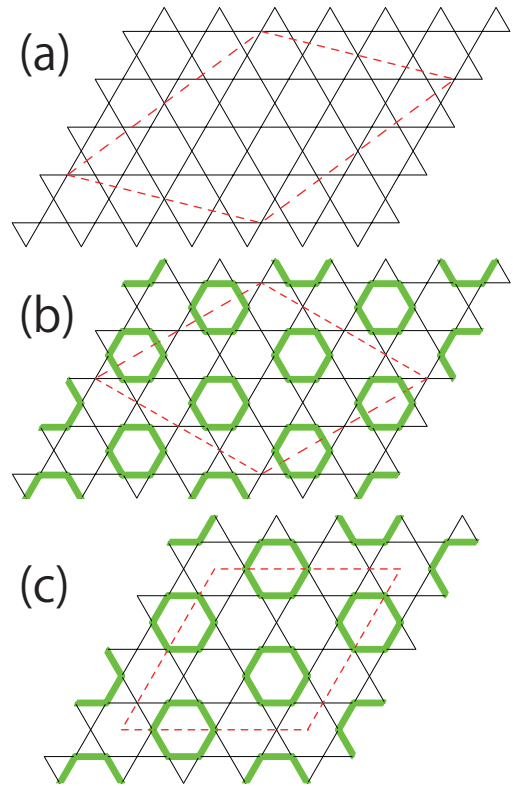}
\end{center}
\caption{(Color) 
Finite-size clusters of the kagome-lattice antiferromagnet 
with and without the $\sqrt{3}\times\sqrt{3}$ distortion. 
In (a), a cluster with $N_{\rm s}=42$ on the undistorted kagome lattice 
is illustrated by the parallelogram of red broken lines. 
The $\sqrt{3}\times\sqrt{3}$ distorted kagome lattice 
is shown in (b) and (c) by the green thick lines 
and black thin lines. 
The finite-size clusters of $N_{\rm s}=36$ and $N_{\rm s}=27$ 
are presented by the rhombus of red broken lines 
in (b) and (c), respectively. 
}
\label{fig1}
\end{figure}

The Hamiltonian that we study in this research is given by 
${\cal H}={\cal H}_0 + {\cal H}_{\rm Zeeman}$, where 
\begin{equation}
{\cal H}_0 = \sum_{\langle i,j\rangle} J 
\mbox{\boldmath $S$}_{i}\cdot\mbox{\boldmath $S$}_{j} , 
\label{H_undistorted_kagome}
\end{equation}
for the model on the undistorted kagome lattice.  
Particularly, we examine 
a cluster with 42 sites of spins shown in Fig.~\ref{fig1}(a).  
We also study 
\begin{equation}
{\cal H}_0 = \sum_{\langle i,j\rangle \in {\rm black \ bonds}} 
J_{\rm 1} 
\mbox{\boldmath $S$}_{i}\cdot\mbox{\boldmath $S$}_{j} 
+
\sum_{\langle i,j\rangle \in {\rm green \ bonds}} J_{\rm 2} 
\mbox{\boldmath $S$}_{i}\cdot\mbox{\boldmath $S$}_{j} 
, 
\label{H_distorted_kagome}
\end{equation}
for the model on the distorted kagome lattice 
shown in Figs.~\ref{fig1}(b) and \ref{fig1}(c).  
Hereafter, 
vertices of thick green hexagons are denoted by $\alpha$ sites 
and 
other vertices are denoted by $\beta$ sites. 
Here, ${\cal H}_{\rm Zeeman} $ is given by 
\begin{equation}
{\cal H}_{\rm Zeeman} = - h \sum_{j} S_{j}^{z} .  
\label{H_zeeman}
\end{equation}
Here, $\mbox{\boldmath $S$}_{i}$ 
denotes the $S=1/2$ spin operator 
at site $i$ illustrated 
by the vertices of the undistorted and distorted kagome lattices. 
The sum of ${\cal H}_0$ runs over all the pairs 
of spin sites linked by solid lines in Fig.~\ref{fig1}.  
Energies are measured in units of $J$ 
for the undistorted kagome lattice 
and $J_{\rm 1}$ for the distorted kagome lattice; 
hereafter, we set $J=1$ and $J_{\rm 1}=1$.  
The number of spin sites is denoted by $N_{\rm s}$. 
We impose the periodic boundary condition 
for clusters with site $N_{\rm s}$. 
Note here that the case of $J_2/J_1=1$ in the Hamiltonian 
(\ref{H_distorted_kagome}) is reduced to 
the Hamiltonian (\ref{H_undistorted_kagome}). 

We calculate the lowest energy of ${\cal H}_0$ 
in the subspace characterized by $\sum _j S_j^z=M$ 
by numerical diagonalizations 
based on the Lanczos algorithm and/or the Householder algorithm. 
The energy is denoted by $E(N_{\rm s},M)$, 
where $M$ takes an integer from zero to the saturation value 
$M_{\rm s}$ ($=S N_{\rm s}$). 
We often use the normalized magnetization $m=M/M_{\rm s}$. 
To achieve calculations of large clusters, 
particularly the case of $N_{\rm s}=42$,  
some of Lanczos diagonalizations have been carried out 
using the MPI-parallelized code, which was originally 
developed in the study of Haldane gaps\cite{HN_Terai}. 
The usefulness of our program was confirmed in large-scale 
parallelized calculations\cite{kgm_gap,s1tri_LRO}. 

The magnetization process for a finite-size system is obtained 
by the magnetization increase from $M$ to $M+1$ at the field 
\begin{equation}
h=E(N_{\rm s},M+1)-E(N_{\rm s},M),
\label{field_at_M}
\end{equation}
under the condition that the lowest-energy state 
with the magnetization $M$ and that with $M+1$ 
become the ground state in specific magnetic fields.  
However, it often happens that 
the lowest-energy state with the magnetization $M$ 
does not become the ground state in any field. 
In this case, the magnetization process 
around the magnetization $M$ is determined 
by the Maxwell construction\cite{kohno_aniso2D,sakai_aniso}. 

The critical exponent $\delta$ is a good index 
for characterizing the universality class 
of the field-induced phase transitions. 
To estimate this $\delta$ 
just outside a specified $m$ in the magnetization process, 
we use the finite-size scaling developed in Ref.~\ref{TSakai_Mtaka}. 
Although this method was originally proposed for one-dimensional cases, 
the validity for two-dimensional cases has been confirmed, 
for example, in the triangular-lattice 
antiferromagnet\cite{Sakai_HN_PRBR}. 
We first assume the asymptotic form 
of the system size dependence of the energy to be
\begin{equation}
\frac{1}{N_{\rm s}} E(N_{\rm s}, M) \sim 
\epsilon(m) + C(m) \frac{1}{N_{\rm s}^{\theta}}, \ 
(N_{\rm s}\rightarrow \infty), 
\end{equation}
where $\epsilon(m)$ is the energy per site. 
The second term means the leading correction 
with respect to $N_{\rm s}$. 
We also assume that $C(m)$ is an analytic function of $m$. 
In this paper, we focus our attention 
on the case of $m=1/3$. 
Thus, the exponents that we want to know are 
$\delta_{\pm}$ defined in the form of 
\begin{equation}
\left|
m - \frac{1}{3}
\right|
\sim
\left|
h - h_{\rm c \pm}
\right|^{1/\delta_{\pm}} , 
\end{equation}
where 
the critical fields are defined as 
\begin{equation}
h_{\rm c \pm} =
\pm \lim_{N_{\rm s}\rightarrow\infty}
\left[
E\left(
N_{\rm s}, \frac{M_{\rm s}}{3} \pm 1
\right)
-
E\left(
N_{\rm s}, \frac{M_{\rm s}}{3}
\right)
\right]. 
\label{critical_fields}
\end{equation}
If we define the quantity $f_{\sigma}(N_{\rm s})$ by 
\begin{eqnarray}
f_{\pm} (N_{\rm s}) &=& 
E
\left(
N_{\rm s}, \frac{M_s}{3}\pm 2
\right)
+
E
\left(
N_{\rm s}, \frac{M_s}{3}
\right) 
\nonumber \\
& &
-
2E
\left(
N_{\rm s}, \frac{M_s}{3}\pm 1
\right) ,
\end{eqnarray}
the asymptotic forms of $f_{\sigma}(N_{\rm s})$ 
are expected to be
\begin{equation}
f_{\pm} (N_{\rm s}) \sim \frac{1}{N_{\rm s}^{\delta_{\pm}}} 
+ O 
\left(
\frac{1}{N_{\rm s}^{\theta+1}}
\right), 
\end{equation}
where $N_{\rm s}\rightarrow\infty$ 
as long as we assume Eq.~(\ref{critical_fields}). 
Therefore, 
it is possible to estimate the exponents $\delta_{+}$ 
and $\delta_{-}$ 
from the gradient of the linear fitting 
in the $\ln (f_{\sigma}) $-$\ln (N_{\rm s})$ plot 
when the condition $\theta +1 > \delta_{\sigma}$ holds. 
In this study, we carry out 
our analysis of $\delta_{\sigma}$ 
assuming this condition 
because the assumption is reasonable 
from successful estimates of the exponents 
in two- and one-dimensional 
systems\cite{TSakai_Mtaka,Sakai_HN_PRBR}, 
which are consistent with 
the relationship between the parabolic dispersion and 
the critical exponent 
depending on the spatial dimension. 

\section{Results and Discussion} 

\subsection{Case of undistorted kagome lattice} 

\begin{figure}[htb]
\begin{center}
\includegraphics[width=7.0cm]{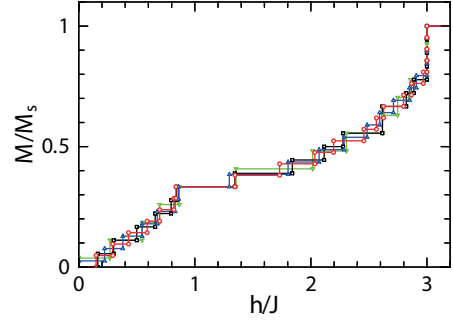}
\end{center}
\caption{(Color) 
Magnetization process 
for the undistorted kagome-lattice antiferromagnet. 
The results of finite-size clusters for $N_{\rm s}=42$, 39, 36, and 27 
are illustrated by red circles, blue triangles, black squares, and 
green reversed triangles, respectively. 
}
\label{fig2}
\end{figure}
Let us first observe the magnetization process 
of the $N_{\rm s}=42$ clusters in the undistorted case. 
The result is shown in Fig.~\ref{fig2} 
together with the magnetization process 
for the $N_{\rm s}=39$ cluster, 
that for  the $N_{\rm s}=36$ cluster, 
and that for  the $N_{\rm s}=27$ cluster. 
Note here that although the clusters 
for $N_{\rm s}=27$, 36, and 39 
are rhombic, the $N_{\rm s}=42$ cluster is not rhombic. 
Even in such a situation of the anisotropy 
in a two-dimensional lattice, 
it is sufficiently worth 
examining the result of a size 
that has not been reached in previous studies. 
In particular, the $N_{\rm s}=42$ cluster suits 
the investigation of the behavior at approximately $m=1/3$, 
although it does not suit the study 
of the behavior at $m=1/9$, $5/9$, and $7/9$ 
because $N_{\rm s}/9$ is not an integer. 
The width of the $N_{\rm s}=42$ step 
at $m=1/3$, namely, $M=\frac{1}{3} M_{\rm s}$, seems large. 
The width at $M=\frac{1}{3} M_{\rm s} -1$ 
is quite small. 
On the other hand, the width at $M=\frac{1}{3} M_{\rm s} +1$ 
is large even if one compares it with the width 
at $M=\frac{1}{3} M_{\rm s}$. 
These features are common with the clusters 
of $N_{\rm s}=39$, 36, and 27; 
one finds that the features do not depend on the system size. 

\begin{figure}[htb]
\begin{center}
\includegraphics[width=7.0cm]{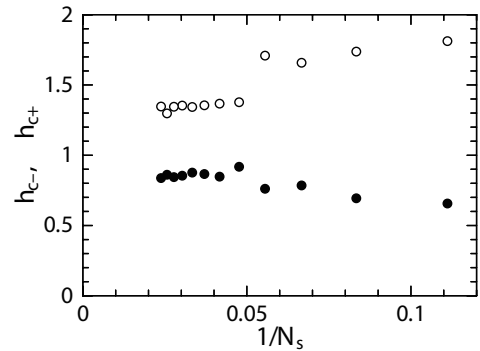}
\end{center}
\caption{
System size dependence of the position 
of the edges at the height of $m=1/3$ in the magnetization process 
for the undistorted kagome-lattice antiferromagnet. 
}
\label{fig3}
\end{figure}
To examine the position of the edges 
of the state at $m=1/3$ in more detail, 
we plot its system size dependence as a function of $1/N_{\rm s}$; 
the result is shown in Fig.~\ref{fig3}. 
This figure was originally presented 
as Fig.~4 in Ref.~\ref{Hida_kagome}; 
however, the plotted data were limited 
to cases up to $N_{\rm s} = 33$. 
We additionally plotted the results for larger clusters 
$N_{\rm s}=36$, 39, and 42. 
The new datum of $h_{\rm c-}$ for $N_{\rm s}=42$ 
is quite close to the data for a smaller $N_{\rm s}$. 
The situation is the same as that for $h_{\rm c+}$. 
The data for $N_{\rm s} \ge 21$ seem almost independent 
of $N_{\rm s}$ and seem to converge to different values 
with each other. 
In this sense, our new data for $N_{\rm s}=42$ 
are not the results which suggests that 
the width $h_{\rm c+} - h_{\rm c-}$ decays and vanishes 
in the thermodynamic limit. 
However, there certainly exists 
a discontinuous size dependence between $N_{\rm s}=18$ and 21. 
The present new data for $N_{\rm s}=42$ cannot guarantee that 
a similar discontinuous behavior never happens for $N_{\rm s} > 42$. 
It may be premature 
to conclude from the numerical-diagonalization data 
whether the width at $m=1/3$ survives or vanishes 
in the thermodynamic limit. 
Another important feature is that,  
at least for $N_{\rm s}\ge 21$, 
the size dependence of data in the case 
when $N_{\rm s}/9$ is an integer is in 
agreement with that in the case 
when $N_{\rm s}/9$ is not an integer.  
If the quantum state at $m=1/3$ forms a nine-site structure, 
the state becomes stable from the viewpoint of its energy. 
In the case when $N_{\rm s}/9$ is not an integer, 
on the other hand, the nine-site structure is partly realized 
in finite-size clusters; 
the energies per site may be larger than 
those in the case when $N_{\rm s}/9$ is an integer. 
The consequence would lead to the appearance of 
a difference in data in Fig.~\ref{fig3} between 
whether $N_{\rm s}/9$ is an integer and is not. 
However, Fig.~\ref{fig3} does not show such a difference. 
Thus, it is not reasonable to consider that 
the state at $m=1/3$ shows the nine-site structure 
as a long-range order, 
although a feature of the nine-site structure 
may survive as a short-range correlation. 

\begin{figure}[htb]
\begin{center}
\includegraphics[width=7.0cm]{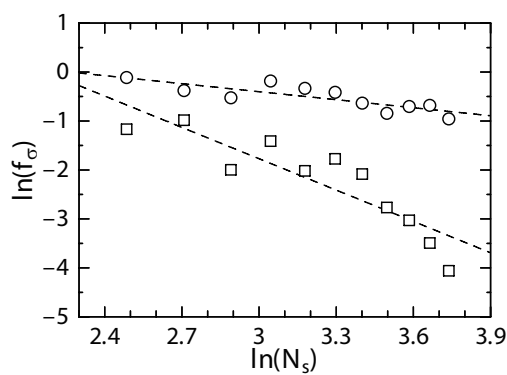}
\end{center}
\caption{$\ln (f_{\sigma})$ is plotted vs $\ln (N_{\rm s})$ 
for the undistorted kagome-lattice antiferromagnet. 
Circles and squares denote the results for $f_{+}$ and  $f_{-}$, 
respectively.  
The broken lines are the results of the linear fitting 
in this plot. 
}
\label{fig4}
\end{figure}
Next, let us examine the characteristics of the behaviors 
just outside $m=1/3$ in the magnetization process. 
According to the argument explained above, we plot 
$\ln (f_{\sigma})$ versus $\ln (N_{\rm s})$ for $N_{\rm s} \ge 12$; 
the result is shown in Fig.~\ref{fig4}. 
We plot not only data of rhombic clusters 
for $N_{\rm s}=12$, 21, 27, 36, and 39, 
but also data of parallelogram clusters
for $N_{\rm s}=15$, 18, 24, 30, 33, and 42. 
Parallelogram clusters for $N_{\rm s}=15$, 18, 24, 30, and 33 
are the same as those in Ref.~\ref{Hida_kagome}. 
Figure~\ref{fig4} suggests that the calculated points 
can be fitted to a line for $\ln (f_{+})$. 
The standard least-squares fitting for $\ln (f_{+})$ gives 
\begin{equation}
\delta_{+}=0.54 \pm 0.36 . 
\label{exponent_pos}
\end{equation}
This result is in agreement with the previous estimate shown 
in Ref.~\ref{Sakai_HN_PRBR}, which uses rhombic clusters. 
The present result of $\delta_{+}$ is clearly in disagreement 
with that of $\delta=1$ which is observed widely 
in various two-dimensional systems as a typical behavior, 
for example, 
in the triangular-lattice antiferromagnet\cite{Sakai_HN_PRBR}. 
Note here that 
exponent (\ref{exponent_pos}) indicates that 
there is no discontinuity in the gradient 
between the states of $m=1/3$ and $m>1/3$. 
For $\ln (f_{-})$, on the other hand, 
data for small $N_{\rm s}$ show deviations. 
If we perform the standard least-squares fitting 
by a linear line for all the data for $N_{\rm s}=12$-42, 
we obtain 
\begin{equation}
\delta_{-}=2.13 \pm 1.10 , 
\label{exponent_neg}
\end{equation}
which is in agreement with the estimate shown 
in Ref.~\ref{Sakai_HN_PRBR}. 
This estimate is also different from the standard value 
of $\delta=1$ in two-dimensional systems. 
However, data for large $N_{\rm s}$ seem to show 
a steeper dependence corresponding to $\delta_{-}$ 
which is larger than Eq.~(\ref{exponent_neg}). 
This large gradient possibly suggests 
that a first-order transition occurs at $h=h_{\rm c-}$ 
although a discontinuous behavior is not detected 
in this study. 
To confirm the first-order transition, 
it is necessary to carry out further investigations 
of the magnetization process based on even larger systems. 
Our results of exponents (\ref{exponent_pos}) 
and (\ref{exponent_neg}) should be compared with 
other estimates from a different approach. 
A possible candidate approach 
is the grand canonical analysis 
of DMRG results\cite{SNishimoto_NShibata_CHotta} 
because the authors of Ref.~\ref{SNishimoto_NShibata_CHotta} 
insist that this method successfully gives 
bulk-limit quantities free from the boundary effect; 
the comparison of our results 
with the results from this method is an urgent issue. 

\subsection{Case of the $\sqrt{3}\times\sqrt{3}$-distorted kagome kattice} 

In this subsection, we examine how the behavior changes 
when the kagome lattice shows a distortion of 
the $\sqrt{3}\times\sqrt{3}$ type. 
This distortion in the kagome-lattice antiferromagnet
was originally investigated in Ref.~\ref{Hida_kagome}, 
in which a peculiar backbending behavior in the magnetization process 
was reported at the higher-field edge of the one-third height of
the saturation at approximately $J_2/J_1\sim 1.25$.  
Since the system sizes were, unfortunately, limited 
to being very small at that time, 
it was unclear whether the behavior is an artifact 
due the finite-size effect or a truly thermodynamic behavior. 
Recently, investigations based on the numerical-diagonalization results 
of a larger system\cite{HN_spin_flop} have clarified that 
the behavior certainly exists in a larger system; 
Ref.~\ref{HN_spin_flop} showed that 
this behavior is related to the occurrence of the spin-flop phenomenon 
even when the system is isotropic in spin space. 
\begin{figure}[htb]
\begin{center}
\includegraphics[width=7.0cm]{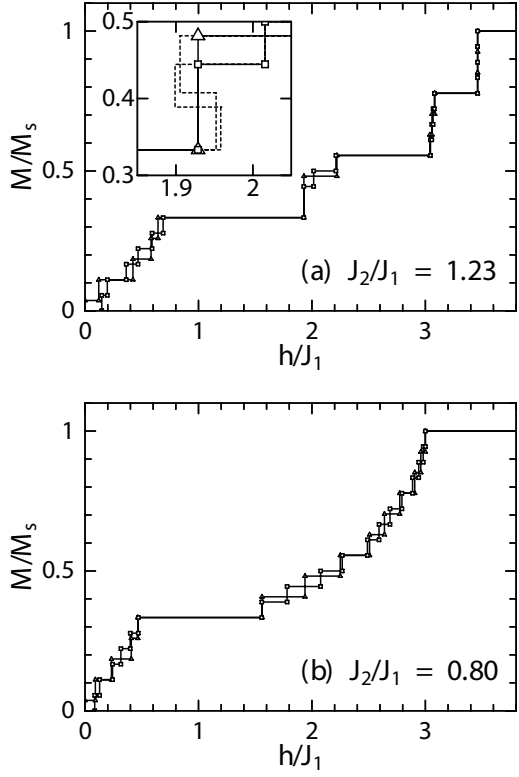}
\end{center}
\caption{Magnetization process for the $S =1/2$ Heisenberg
antiferromagnet on the $\sqrt{3}\times\sqrt{3}$ distorted 
kagome lattice. Panels (a) and (b) show 
the cases of $J_{2}/J_{1}$=1.23 and 0.80, respectively.  
Squares and triangles are the results for $N_{\rm s}=36$ and 27, 
respectively. 
The inset in (a) shows a zoomed-in view at the higher-field 
edge of the one-third height of the saturation, 
where the broken lines represent the results
before the Maxwell construction is carried out.
}
\label{fig5}
\end{figure}
The same spin-flop phenomenon is reported 
in the square-kagome lattice\cite{shuriken_lett} and 
{\it shuriken}-bonded honeycomb lattice\cite{HN_spin_flop}. 
Figure~\ref{fig5}~(a) shows the same behavior at $J_2/J_1=1.23$.  
However, the investigation reported in Ref.~\ref{HN_spin_flop} 
focused on the case of $J_2/J_1 > 1$. 
To study the behavior at around 
the undistorted point $J_2/J_1=1$, 
the present study treats not only the case of $J_2/J_1 > 1$  
but also the case of $J_2/J_1 < 1$. 

We show the magnetization process for $J_{2}/J_{1}=0.80$ 
in Fig.~\ref{fig5}~(b). 
One easily finds that several features are different 
between the cases of $J_2/J_1$=1.23 and 0.80. 
Contrary to the magnetization plateaux 
at $m=5/9$ and 7/9 observed in the case of $J_2/J_1$=1.23, 
the widths at these heights in the case of $J_2/J_1$=0.80 
become markedly smaller, 
suggesting the disappearance of the plateaux. 
The magnetization jump between $m=7/9$ and the saturation 
is observed in the case of $J_2/J_1$=1.23 owing to 
the formation of explicit eigenstates 
with a spatially localized structure 
at hexagons on the kagome lattice. 
In the case of $J_2/J_1$=0.80, 
the jump also disappears. 
On the other hand, 
the behavior at $m=1/3$ is similar 
between the cases of $J_2/J_1$=1.23 and 0.80; 
the state of $m=1/3$ is realized in 
a wide region of external field, 
suggesting the existence of the magnetic plateau. 

\begin{figure}[htb]
\begin{center}
\includegraphics[width=7.0cm]{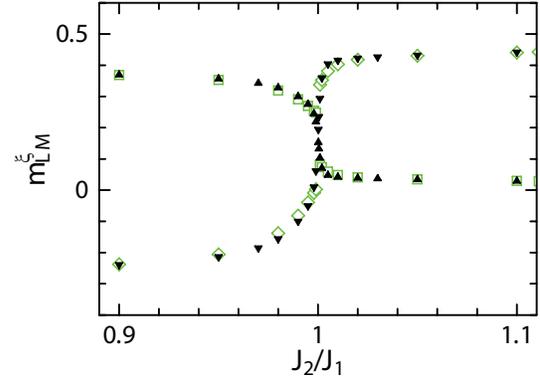}
\end{center}
\caption{(Color) Dependence of the local magnetization 
on the ratio of interaction $J_2/J_1$. 
For $N_{\rm s}=27$, closed triangles and closed inversed triangles 
denote the results of $\xi=\alpha$ and $\beta$, respectively. 
For $N_{\rm s}=36$, green open squares and green open diamonds
denote the results for $\xi=\alpha$ and $\beta$, respectively. 
}
\label{fig6}
\end{figure}
To examine whether 
the properties of the $m=1/3$ states 
for $J_2/J_1$=1.23 and 0.80 are the same or different, 
we evaluate the local magnetization defined as 
\begin{equation}
m_{\rm LM}^{\xi} = \frac{1}{N_{\xi}} 
\sum_{j\in \xi} \langle S_j^{z} \rangle , 
\label{ave_local_mag}
\end{equation}
where $\xi$ takes $\alpha$ and $\beta$. 
Here, the symbol $\langle {\cal O} \rangle$ 
denotes the expectation value 
of the operator ${\cal O}$ with respect 
to the lowest-energy state 
within the subspace with a fixed $M$ of interest. 
Recall here that 
the case of interest in this paper is $M=M_{\rm s}/3$. 
Here $N_{\xi}$ denotes the number of $\xi$ sites; 
averaging over $\xi$ is carried out 
in the case when the ground-state level is degenerate. 
Note that, for $M$ with nondegenerate ground states, 
the results do not change regardless of 
the presence or absence of this average. 
Our results of $m_{\rm LM}^{\xi}$ are shown in Fig.~\ref{fig6}. 
One clearly observes a large increase in $m_{\rm LM}^{\beta}$ 
and a large decrease in $m_{\rm LM}^{\alpha}$ 
at approximately $J_2/J_1 =1$. 
In the region of $J_2/J_1 > 1$, 
an $\alpha$-site spin becomes almost vanishing, 
while a $\beta$-site spin becomes an up-spin. 
This spin state has already been discussed in Ref.~\ref{HN_spin_flop}:   
the state is composed of 
a spin-singlet at two neighboring $\alpha$ spins and 
an up-spin at a $\beta$ site 
in each local triangle of the lattice. 
In the region of $J_2/J_1 < 1$, on the other hand, 
an $\alpha$-site spin becomes an up-spin 
while a $\beta$-site spin becomes a down-spin. 
This spin state is easily understood 
if one considers the case of $J_2/J_1=0$. 
In this limiting case, 
the Marshall-Lieb-Mattis theorem\cite{Marshall,Lieb_Mattis} holds. 
Therefore, the ferrimagnetic state 
with the up-spin at the $\alpha$ site and 
the down-spin at the $\beta$ site 
is realized as the ground state with a spontaneous magnetization 
even in the absence of a magnetic field. 
The result in Fig.~\ref{fig6} suggests that, 
in the region of $J_2/J_1 < 1$ near $J_2/J_1 = 1$, 
the same ferrimagnetic state appears under some external field. 
It is an unresolved issue 
where the phase transition 
to such a Lieb-Mattis-type ferrimagnetic state 
as the ground state with a spontaneous magnetization occurs. 
A related problem was investigated 
in the spatially anisotropic kagome-lattice antiferromagnet 
in Refs.~\ref{collapse_ferri} and \ref{Shimokawa_JPSJ}, 
which pointed out the existence of the 
non-Lieb-Mattis phase in the intermediate region 
between the Lieb-Mattis-type ferrimagnetic phase and 
the nonmagnetic phase. 
Note here that this intermediate ferrimagnetic state is 
observed in various one-dimensional systems\cite{Hida_nLM,
Hida_Takano,Shimokawa-nLM-spin-one-half,Shimokawa-nLM-spin-one}. 
It is also an unresolved issue whether such an intermediate phase 
is present or absent in the present distortion 
of the $\sqrt{3}\times\sqrt{3}$ type. 
The most important consequence is a marked change 
in $m_{\rm LM}^{\alpha}$ and $m_{\rm LM}^{\beta}$ at $J_2/J_1 =1$, 
suggesting the occurrence of a quantum phase transition. 
In Refs.~\ref{HNakano_Cairo_lt} and \ref{Isoda_Cairo_full}, 
a similar observation about the local magnetization was reported 
in the $S=1/2$ Heisenberg antiferromagnet on the Cairo-pentagon 
lattice\cite{Ressouche_Cairo,Rousochatzakis_Cairo}. 
One difference is that the boundary is independent of $N_{\rm s}$ 
and is $J_2/J_1=1$ for both $N_{\rm s}=27$ and 36. 
Unfortunately, it is 
difficult to conclude whether the transition is of the second or 
first order because the change in  $m_{\rm LM}^{\xi}$ 
is continuous 
for $N_{\rm s}=36$, while it is discontinuous for $N_{\rm s}=27$ 
from examinations based on the numerical-diagonalization data. 

\begin{figure}[htb]
\begin{center}
\includegraphics[width=7.0cm]{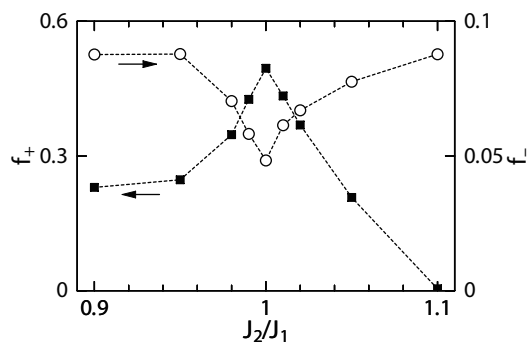}
\end{center}
\caption{
Dependences of $f_{+}(N_{\rm s})$ and $f_{-}(N_{\rm s})$ for $N_{\rm s}=36$ 
on the ratio of interaction $J_2/J_1$ 
around the undistorted point. 
Closed squares and open circles denote 
$f_{+}(N_{\rm s})$ and $f_{-}(N_{\rm s})$, respectively. 
}
\label{fig7}
\end{figure}
Finally, let us observe the change at approximately $J_2/J_1=1$ 
just outside $m=1/3$ in the magnetization process. 
To do so, we examine 
the dependences of $f_{+}(N_{\rm s})$ and $f_{-}(N_{\rm s})$ 
on the ratio $J_2/J_1$; 
the result for $N_{\rm s}=36$ is shown in Fig.~\ref{fig7}. 
One easily observes different dependences on $J_2/J_1$ 
between the regions of $J_2/J_1 > 1$ and $J_2/J_1 <1$.   
This difference appears 
regardless of $f_{+}(N_{\rm s})$ and $f_{-}(N_{\rm s})$. 
This result strongly suggests 
the existence of some boundary at $J_2/J_1 =1$.  
One finds that 
the behavior of $f_{+}(N_{\rm s})$ at $J_2/J_1 =1$ is maximum, 
while that of $f_{-}(N_{\rm s})$ at $J_2/J_1 =1$ is minimum. 
Although the reason for this contrast is currently unresolved, 
the undistorted point is certainly a boundary. 

\section{Conclusions} 
%% No sections necessary for express letters, letters and short notes

We have studied the Heisenberg antiferromagnet 
on the kagome lattice without and with a distortion 
of the $\sqrt{3}\times\sqrt{3}$ type 
by the numerical-diagonalization method. 
We present a new result of the magnetization process 
for this undistorted model for a 42-site cluster 
for the first time. 
Our diagonalization results suggest that 
the critical exponents of the field dependence 
of the magnetization 
just outside of the one-third magnetization state 
are markedly different from the exponent $\delta=1$ 
for the typical magnetization plateau 
in a two-dimensional system. 
On the other hand, 
we do not obtain positive evidence 
of the vanishing width at the one-third magnetization state 
even if we take into account our new result of the 42-site cluster. 
Our results of the distorted model 
clearly indicate that the undistorted kagome point 
is a boundary between two regions: 
the ferrimagnetic state and the state with the nine-site 
structure including a hexagonal singlet and three up-spins. 
The change near the undistorted point as the transition point 
is very rapid. 
It is unclear whether the phase transition is continuous or discontinuous 
at the present stage. 
At such a transition point, the anomalous critical exponents 
are not so unnatural. 

In Ref.~\ref{SNishimoto_NShibata_CHotta},  
the authors proposed that 
the one-third magnetization state is well described 
by the state with the nine-site structure 
including a hexagonal singlet and three up-spins.
This picture of the state is in disagreement with the present consequence 
that the undistorted point is just a boundary. 
A possible reason for this discrepancy 
is that 
the deformation used in Ref.~\ref{SNishimoto_NShibata_CHotta} 
plays an essential role as a perturbation 
added to the ideal kagome-lattice antiferromagnet 
and that  
the case 
that the calculation in Ref.~\ref{SNishimoto_NShibata_CHotta} captures 
may, therefore, not be only the ideal kagome-lattice antiferromagnet 
but is a case that deviated slightly from the ideal point, 
which is the transition point. 
Even though the deviation is taken to be very small, 
a difficulty is possibly unavoidable 
in capturing the true behavior only on the transition point. 
The present study of the kagome-lattice antiferromagnet 
suggests that 
a method based on a deformation technique possibly reaches 
an incorrect conclusion 
concerning subtle behaviors near the transition point. 

Another distortion introduced into the kagome lattice 
was studied from the viewpoint of continuously linking 
the kagome and triangular lattices\cite{HN_Sakai_PSS,HN_Sakai_SCES}. 
Concerning the $m=1/3$ state, 
Refs.~\ref{HN_Sakai_PSS} and \ref{HN_Sakai_SCES} 
showed 
that the phase transition occurs 
between the kagome and triangular points 
and that the kagome point is close to but 
different from the transition point. 
Note here that this type of distortion was realized 
in an experimental study\cite{Ishikawa2014}; 
the direct comparison is difficult at the present time 
because the spin is larger than $S=1/2$. 
If an experimental realization of an $S=1/2$ system 
is successful, the comparison 
between the experiments and theoretical predictions 
would be useful for our deeper understanding 
of quantum spin systems. 
Studies of other types of distortion would also 
contribute to the progress of this field.

\section*{Acknowledgments}
We wish to thank 
Professors S.~Miyashita and 
Dr. N. Todoroki 
for fruitful discussions. 
This work was partly supported 
by JSPS KAKENHI Grant Numbers 23340109 and 24540348. 
Nonhybrid thread-parallel calculations
in numerical diagonalizations were based on TITPACK version 2
coded by H. Nishimori. 
This research used computational resources of the K computer 
provided by the RIKEN Advanced Institute for Computational Science 
through the HPCI System Research projects 
(Project IDs: hp130070 and hp130098). 
Some of the computations were 
performed using facilities of 
the Department of Simulation Science, 
National Institute for Fusion Science; 
Center for Computational Materials Science, 
Institute for Materials Research, Tohoku University; 
Supercomputer Center, 
Institute for Solid State Physics, The University of Tokyo;  
and Supercomputing Division, 
Information Technology Center, The University of Tokyo. 
This work was partly supported 
by the Strategic Programs for Innovative Research; 
the Ministry of Education, Culture, Sports, Science 
and Technology of Japan; 
and the Computational Materials Science Initiative, Japan. 
We also would like to express our sincere thanks 
to the staff of the Center for Computational Materials Science 
of the Institute for Materials Research, Tohoku University, 
for their continuous support 
of the SR16000 supercomputing facilities. 

%\appendix

\end{document}